\documentclass[12pt,reqno]{amsart}
\usepackage{amssymb}

\newcommand\lp{\left(}
\newcommand\rp{\right)}

\newcommand\fsl{\mathfrak{sl}}
\newcommand\fgl{\mathfrak{gl}}
\newcommand\gal{\mathfrak{g}}
\newcommand\fal{\mathfrak{f}}
\newcommand\hal{\mathfrak{h}}
\newcommand\sal{\mathfrak{s}}
\newcommand\bal{\mathfrak{b}}
\newcommand\nal{\mathfrak{n}}
\newcommand\calg{\mathfrak{c}}
\newcommand\jal{\mathfrak{j}}

\newcommand\cnums{\mathbb{C}}
\newcommand\intnums{\mathbb{Z}}
\newcommand\natnums{\mathbb{N}}
\newcommand\rnums{\mathbb{R}}
\newcommand\tnums{\mathbb{T}}

\newcommand\Gsp{\mathbf{G}}
\newcommand\Hsp{\mathbf{H}}
\newcommand\Msp{\mathbf{M}}

\newcommand\cfty{{\mathcal C}^{\raise1pt\hbox{$\scriptscriptstyle
      \infty$}}\!}
\newcommand\comega{{\mathcal C}^{\raise1pt\hbox{$\scriptscriptstyle
      \omega$}}\!}

\newcommand\supl{^{\scriptscriptstyle \mathrm L}}
\newcommand\supr{^{\scriptscriptstyle \mathrm R}}
\newcommand\supp{^{\scriptscriptstyle \pi}}

\newcommand\subH{_{\scriptscriptstyle \Hsp}}

\newcommand\ap{a\supp}

\newcommand\E{\mathrm e}
\newcommand\rH{\mathrm H}
\newcommand\rZ{\mathrm Z}

\newcommand\ro{\mathrm o}

\newcommand\ri{\mathrm i}

\newcommand\rSL{\mathrm{SL}}

\newcommand\Mo{{\Msp_\ro}}

\newcommand\heta{{\hat{\eta}}}

\newcommand\hb{{\hat{b}}}
\newcommand\hk{{\hat{k}}}

\DeclareMathOperator{\ad}{ad}

\DeclareMathOperator{\Hom}{Hom}

\newtheorem{theorem}{Theorem}[section]
\newtheorem{proposition}[theorem]{Proposition}
\newtheorem{lemma}[theorem]{Lemma}
\newtheorem{conjecture}[theorem]{Conjecture}
\newtheorem{definition}[theorem]{Definition}
\theoremstyle{definition}
\newtheorem{example}[theorem]{Example}

\title{Quantization of cohomology in semi-simple Lie algebras}
\subjclass{22E41 22E45}
\author[R. Milson]{R. Milson\dag}
\thanks{\dag\ Research supported by an NSERC Canada post-doctoral fellowship.}
\address{McGill University}
\email{milson@math.mcgill.ca}
\author{D. Richter}
\address{University of Minnesota}
\email{drichter@math.umn.edu}

\begin{document}
\begin{abstract}
  Let $\gal$ be a complex, finite-dimensional Lie algebra, and $\Mo$ a
  contractable neighborhood of a complex homogeneous space on which
  $\gal$ acts transitively.  The present article investigates
  quantization of $\rH^1(\gal;\comega(\Mo)/\cnums)$ by the condition
  that the corresponding realization by first-order operators admits a
  finite-dimensional invariant subspace of functions.  The quantization of
  cohomology phenomenon surfaced during the recent classification of
  finite-dimensional Lie algebras of first and zero order operators in
  two complex variables; this classification revealed that
  quantization holds for all two-dimensional homogeneous spaces.
  
  The present article presents the first known counter-examples to
  quantization of cohomology; it is shown that quantization can fail
  even if $\gal$ is semi-simple, and even if the homogeneous space in
  question is compact.
  
  A explanation for the quantization phenomenon is given in the case
  of semi-simple $\gal$.  It is shown that the set of classes in
  $\rH^1$ that admit finite-dimensional invariant subspaces is a
  semigroup that lies inside a finitely-generated abelian group.  In
  order for this abelian group be a discrete subset of $\rH^1$, i.e.
  in order for quantization to take place, some extra conditions on
  the isotropy subalgebra are required.  Two different instances of
  such necessary conditions are presented.
\end{abstract}
\maketitle
\section{Introduction}
The present article deals with realizations of finite-dimensional Lie
algebras by first-order differential operators on homogeneous spaces.
Differenential realizations are widely applied to the problems of
quantum mechanics as well as being interesting from a purely
mathematical point of view.  

Realizations that admit a finite-dimensional invariant subspace of
functions are particularly useful in the context of algebraically
solvable spectral problems \cite{Tu:sl2} \cite{ShTu:mdim}
\cite{GKO:mdim}.  A prototypical example is the following of
realizations of $\fsl_2$
$$\partial_z,\quad 2z \partial_z - \lambda,\quad z^2 \partial_z -
\lambda z,$$
where $\lambda$ is a scalar parameter.  Note that if
$\lambda$ is a natural number then then $1,z,\ldots,z^\lambda $ span
an invariant subspace.  It can be shown \cite{KO:sl2} that
finite-dimensional invariant subspaces do not exist for other values
of the $\lambda$ parameter.  Thus the existence of a
finite-dimensional invariant subspace seems to behave as a kind of a
quantization condition.

More generally one may consider realizations of a finite dimensional
Lie algebra, $\fal$, by first- and zero-order operators,
$$V(a)+\lambda_1\eta_1(a) + \ldots + \lambda_n\eta_n(a),\quad
a\in\fal$$
where $V$ is a vector field (possibly zero), the $\eta$'s
are functions, and the $\lambda_i$ scalar parameters.  Using Sophus
Lie's classification of finite-dimensional Lie algebras of vector
fields in two complex variables \cite{Lie}, Gonzalez-Lopez, Karman,
and Olver were able to describe all such two-dimensional realizations
\cite{GKO}.  Calculating case by case, these authors were able to
confirm that finite-dimensional modules of functions exist only for
certain discrete values of the $\lambda$ parameters.  There is a
natural way, described below, to regard the tuple
$(\lambda_1,\ldots,\lambda_n)$ as an element of a certain $\rH^1$, and
so, the authors of \cite{GKO} named this phenomenon quantization of
cohomology and wondered whether or not it continues to hold for higher
dimensions.

The purpose of the present article is twofold.  First, it will be
shown (Example \ref{ex:counterexample1}) that quantization of
cohomology does not hold in general. Indeed, there exist
counter-examples even when the Lie algebra in question is semi-simple
(Example \ref{ex:counterexample2}), and even when the underlying
homogeneous space is compact (Example \ref{ex:counterexample3}).  The
second aim is to describe some necessary conditions that will assure
that quantization of cohomology takes place.  In regards to this
second objective the present paper will only consider complex
semi-simple Lie algebras. Indeed, it is possible to show that with
this assumption the cohomology classes that correspond to
finite-dimensional invariant subspaces are a subset of a finitely
generated abelian subgroup of $\rH^1$ (Theorem \ref{thrm:fingen}.)
The fact that this abelian subgroup is finitely generated does not
guarantee that it is discrete; additional assumptions about the
isotropy subalgebra are required for discreteness (Theorems
\ref{thrm:ratsubalg} and \ref{thrm:cod1compact}).

The present approach is based on two ideas.  The finite-dimensional
irreducible representations of a complex, semi-simple Lie algebras
correspond to dominant weights that are integral and positive; this is
quantization of sorts.  To make use of this observation, one has to
relate the abstract representation theory to realizations by
differential operators.  This will be accomplished via the technique
of induced representations, and through the use of Frobenius
reciprocity in the context of finite-dimensional representations of
Lie algebras.

The rest of the article is organized as follows.  Section
\ref{sect:qcoho} describes and motivates the notion of quantization of
cohomology.  Section \ref{sect:indrep} recasts the problem in terms of
induced representations.  Section \ref{sect:examples} describes
several instances where quantization of cohomology fails.  Finally,
Section \ref{sect:semisimple} investigates quantization of cohomology
in the semi-simple case, and gives some necessary conditions on the
isotropy algebra that imply quantization.

\section{Quantization of cohomology}
\label{sect:qcoho}
Let $\Mo$ be an open neighborhood of a complex manifold.  A
holomorphic first order differential operator defined on $\Mo$ has the
form $V+\eta$, where the first term is a vector field, and where the
second term is a multiplication operator corresponding to an
$\eta\in\comega(\Mo)$.  Thus, a local realization of a finite
dimensional Lie algebra, $\gal$, by holomorphic, first order
differential operators is specified by two items: a realization of
$\gal$ by vector fields $V(a), a\in\gal$, and a linear map
$\eta:\gal\rightarrow\comega(\Mo)$ with the property that resulting
collection of operators $\{ V(a)+\eta(a),\; a\in\gal\}$ is closed
under the operator bracket.  This is equivalent to the condition that
$$V(a)(\eta(b))-V(b)(\eta(a)) = \eta([a,b]),\quad a,b\in\gal.$$
Equivalently, using the language of Lie algebra cohomology one can say
that $\eta$ must be a 1-cocycle of $\gal$ with coefficients in
$\comega(\Mo)$. 

Given a realization of $\gal$ by differential operators one can obtain
many other realizations by conjugating the given operators by an
arbitrary multiplication operator $e^\sigma,\; \sigma\in\comega(\Mo)$,
i.e. by applying a gauge transformation.  It therefore makes sense to
consider differential realizations up to gauge equivalence.  Note that
for a first order operator $V+\eta$ and a function $\sigma$ one has
$$\E^{-\sigma} (V+\eta) \E^\sigma = V + \eta + V(\sigma).$$
This means
that the effect of a gauge transformation on a realization by first
order operators is to add a coboundary term, $\delta\sigma$, to the
1-cocycle $\eta$. Thus, for a fixed realization by vector fields, and
up to gauge equivalence, realizations of $\gal$ by first order
differential operators are classified by $\rH^1(\gal;\comega(\Mo))$.

More generally consider a realization of a finite-dimensional
Lie-algebra, $\fal$, by holomorphic first and zero order differential
operators.  Such operators have the form $V+\eta$, where the vector
field term is allowed to be zero.  Let $W\subset\fal$ denote the
abelian ideal of zero-order elements, i.e.  those operators for which
$V=0$, and let $\gal$ denote the quotient $\fal/W$.  The projection
$V+\eta\mapsto V$ gives a realization of $\gal$ by vector fields.  In
this situation one can describe $\eta$ as a linear map from $\gal$ to
$\comega(\Mo)/W$, and one can easily verify that closure under the
operator bracket is equivalent to the condition that $\eta$ be a
1-cocycle of $\gal$ with coefficients in $\comega(\Mo)/W$.  It is well
known that central extensions of $\gal$ by $W$ are classified by
$\rH^2(\gal;W)$.  Let $\gal\ltimes_\eta W$ be a central extension
whose class corresponds to the image of $[\eta]$ in the following long
exact sequence of cohomology:
\begin{equation}
  \label{eq:mod.one.les}
  \ldots \rightarrow \rH^1(\gal;\comega(\Mo)) 
\rightarrow \rH^1(\gal;\comega(\Mo)/W) 
\rightarrow \rH^2(\gal;W) \rightarrow \ldots
\end{equation}
Indeed, it is not hard to verify that $\fal$ is isomorphic to such a
central extension.  Thus, a local realization of a finite dimensional
Lie algebra by first and zero order differential operators requires 3
ingredients: a realization of a finite-diemnsional Lie algebra,
$\gal$, by vector fields, a finite dimensional $\gal$-module, $W$,
realized as an invariant subspace of $\comega(\Mo)$, and a 1-cocycle,
$\eta$, of $\gal$, with coefficients in $\comega(\Mo)/W$.  The
realized Lie algebra will be isomorphic to $\gal\ltimes_\eta W$.

As before, a gauge transformation adds a
coboundary term to $\eta$.  Thus, for a fixed realization of $\gal $ by
vector fields, and a fixed $\gal$-module $W\subset\comega(\Mo)$, the
set of all possible realizations by first and zero order operators of
all possible central extensions of $\gal$ by $W$ is specified up to
gauge equivalence by $\rH^1(\gal;\comega(\Mo)/W)$.

Let $\Lambda\subset\rH^1(\gal;\comega(\Mo)/W)$ denote the set of all
cohomology classes, $\eta$, such that the corresponding realization
of $\gal\ltimes_\eta W$ admits a finite-dimensional invariant subspace
of $\comega(\Mo)$.  There is an important observation that greatly
simplifies the determination of $\Lambda$.
\begin{proposition}[Lemma 2 of \cite{GKO}]
If $W$ is neither $0$, nor the module of constants, then  $\Lambda$ is
the empty set.
\end{proposition}
The proof of this proposition is straightforward: repeated
multiplication by a non-constant function generates an
infinite-dimensional vector space of functions.  Thus, without loss of
generality, one need only consider extensions of $\gal$ by $\cnums$,
the module of constants.  The cohomology space of interest is
therefore $\rH^1(\gal;\comega(\Mo)/\cnums)$.
\begin{definition}
  Let $V(a),\; a\in\gal$ be a realization of $\gal$ by vector fields.
  If $\Lambda$, as defined above, is a discrete subset of
  $\rH^1(\gal;\comega(\Mo)/\cnums)$, then quantization of cohomology
  will be said to hold for $V$.
\end{definition}

It is not particularly worthwhile to study quantization of cohomology
for the full class of vector field realizations.  Indeed, the very
definition of quantization of cohomology, as stated above, is
problematic, because in general $\rH^1(\gal;\comega(\Mo)/\cnums)$ is
infinite-dimensional, and thus lacking a canonical topology. This
means that there isn't even a clear notion of what it means for
$\Lambda$ to be discrete.

For this reason the rest of this paper will be based on the assumption
that $\Mo$ is an open subset of a homogeneous space, and that $V$ is
the realization of the corresponding Lie algebra by infinitesimal
automorphisms.  This assumption improves the situation tremendously.
One can show that $\rH^1(\gal;\comega(\Mo)/\cnums)$ is finite
dimensional ( this is a straightforward consequence of Proposition
\ref{prop:cformiso} below) and endow $\rH^1$ with the canonical
vector-space topology.  In this way the notion of discreteness becomes
meaningful.

To this end let $\Gsp$ be a complex Lie group, $\Hsp$ a closed
subgroup, and let $\gal$ and $\hal$ denote the corresponding Lie
algebras.  Let $\Msp=\Gsp/\Hsp$ and $\pi:\Gsp\rightarrow\Msp$ denote,
respectively, the homogeneous space of right cosets, and the canonical
projection.  For $a\in\gal$ let $a\supl$ and $a\supr$ denote,
respectively, the corresponding left- and right- invariant vector
fields on $\Gsp$, and $\gal\supl$ and $\gal\supr$ the collections of
all such.  To avoid any possible confusion, it should be noted that
$\gal\supl$ corresponds to {\em right} group actions, and $\gal\supr$
to {\em left} ones.  Let $\ap=\pi_*(a\supl),\; a\in\gal$ denote the
realization of $\gal$ by projected vector fields (i.e. by
infinitesimal automorphisms).  It will also be assumed that $\hal$
does not contain any ideals of $\gal$.  This will ensure that
$a\mapsto\ap$ is a faithful realization.  Let $\ro=\pi(e)$ denote the
basepoint of $\Msp$. Henceforth let $\Mo$ be a contractable
neighborhood thereof.

One should also note it is essential that the notion of quantization
of cohomology be formulated in terms of
$\rH^1(\gal;\comega(\Mo)/\cnums)$, rather than merely in terms of
$\rH^1(\gal;\comega(\Mo))$.  Indeed, consider the following
realizations of $\fgl(2,\cnums)$:
$$
x\partial_x+\lambda,\; x\partial_y,\; y\partial_x,\;
y\partial_y+\lambda, \quad \lambda\in\cnums.
$$
Taking $x=1,\; y=0$ as the basepoint, and using Proposition
\ref{prop:cformiso} one checks that $\rH^1$ with coefficients in
$\comega(\Mo)$ is one-dimensional, and that $\lambda$ parameterizes
the cohomology classes.  Also note that for all values of $\lambda$
the above operators do not raise the total degree of polynomials, and
hence for all $\lambda$ admit finite-dimensional invariant subspaces.
Quantization of cohomology is not invalidated by this example
precisely because the cocycle of the above realization is trivial when
the cohomology coefficients are taken to be $\comega(\Mo)/\cnums$
rather than merely $\comega(\Mo)$.

A further simplification occurs when $\gal$ is semi-simple.  In this
case $\rH^2(\gal;\cnums)=0$, and consequently all central-extensions
by $\cnums$ are trivial.  From \eqref{eq:mod.one.les} one has,
$\rH^1(\gal;\comega(\Mo)/\cnums)=\rH^1(\gal;\comega(\Mo))$, and hence
all possible realizations of $\gal\ltimes\cnums$ by first and zero
order operators are obtained by adjoining multiplication by constants
to realizations of $\gal$ by strictly first-order operators.
Therefore, in the semi-simple case it suffices to define $\Lambda$ as
the set of those $[\eta]\in\rH^1(\gal;\comega(\Mo))$ such that the
collection of operators $\ap+\eta(a),\;a\in\gal$ admits a
finite-dimensional invariant subspace of functions, and to say that
quantization of cohomology holds if this $\Lambda$ is a discrete
subset of $\rH^1(\gal;\comega(\Mo))$.

As mentioned in the introduction, the study of quantization of
cohomology began when a classification of all finite-dimen\-sional Lie
algebras of first and zero order operators in two complex variables,
and of the corresponding finite-dimensional modules of functions was
obtained in \cite{GKO}.  A case-by-case examination of this
classification revealed the following.
\begin{theorem}
  If $\Msp$ is a $2$-dimensional complex homogeneous space, then
  quantization of cohomology holds for the corresponding realization
  by infinitesimal automorphisms.
\end{theorem}
\noindent
As was mentioned in the introduction, in order for quantization of
cohomology to hold on higher dimensional homogeneous spaces,
additional assumptions are required about the isotropy subalgebra.
This issue will be taken up in Section \ref{sect:semisimple}.

\section{Induced representations}
\label{sect:indrep}
Let $\hal\subset\gal$ be a finite-dimensional Lie algebra, subalgebra
pair. Let $\Gsp$ be the simply connected Lie group corresponding to
$\gal$, and $\Gsp_e\subset\Gsp$ an open, connected subset that
contains the identity element.  Let $U$ be a finite-dimensional
$\hal$-module, and let $\comega(\Gsp_e,U)^\hal$ denote the vector
space of holomorphic, $U$-valued, $\hal\supr$-equivariant functions on
$\Gsp_e$.  In other words, $\phi\in\comega(\Gsp_e,U)^\hal$ if and only
if
$$(a\supr(\phi))(g) = a\cdot(\phi(g)),\quad a\in\hal,\;g\in\Gsp.$$
Since left-invariant and right-invariant vector fields commute,
actions by the former give $\comega(\Gsp_e,U)^\hal$ the structure of a
$\gal$-module.  With this structure, $\comega(\Gsp_e,U)^\hal$ should
be thought of as a representation of $\gal$ induced from $U$.  Note
that if $\Gsp_e$ is all of $\Gsp$, and if the $\hal$-action on $U$
comes from the action of a closed subgroup $\Hsp$, then the above
definition is equivalent to the usual definition of the induced
representation as holomorphic sections of the homogeneous vector
bundle $\Gsp\times\subH U$ \cite{Bott}.  

The obvious drawback of the present definition of induced
representation is that it seems to depend on the choice of $\Gsp_e$.
However, the ``finite-dimensional content'' of the induced
representation is independent of this choice.  This is a consequence
of the following version of Frobenius reciprocity.  Let
$W$ be a finite-dimensional $\gal$-module, and note that $W$ is also
naturally a $\Gsp$-module, because $\Gsp$ is assumed to be simply
connected.
\begin{proposition}
\label{prop:frobrec}
The map from $\Hom_\gal(W,\comega(\Gsp_e,U)^\hal)$ to $\Hom_\hal(W,U)$
given by
$$
\phi\mapsto \phi(-)(e),\quad
\phi\in\Hom_\gal(W,\comega(\Gsp_e,U)^\hal)
$$
is a vector space isomorphism.  For $\alpha\in\Hom_\hal(W,U)$, the
inverse image, $\phi\in\Hom_\gal(W,\comega(\Gsp_e,U)^\hal)$ is given
by
$$\phi(w)(g) = \alpha(g\cdot w),\quad w\in W,\; g\in\Gsp_e.$$
\end{proposition}

The above definition of induced representation is relevant to the
present discussion because, as the following Proposition will show,
the $\gal$-action on $\comega(\Mo)$ coming from a realization of
$\gal$ by first-order operators is really the same thing as a
$\gal$-representation induced from a $1$-dimen\-sional $\hal$-module.

Note that the vector space of all $1$-dimensional representations of
$\hal$ is naturally isomorphic to $\rH^1(\hal;\cnums)$.  For a
character, $\heta$, of a $1$-dimensional representation of $\hal$, let
$\cnums_\heta$ denote the corresponding 1-dimensional module.  For
$\heta\in\rH^1(\hal,\cnums)$ it will be convenient to denote the
corresponding induced representation,
$\comega(\Gsp_e,\cnums_\heta)^\hal$, simply as
$\comega(\Gsp_e)^\heta$; the latter is just the vector space of
scalar-valued functions, $\phi$, that satisfy $a\supr(\phi) =
\heta(a)\phi,\;a\in\hal$.  Given an open, contractable
$\Mo\subset\Msp$, choose $\Gsp_e\subset\Gsp$ such that there exists a
trivialization $\Gsp_e=\Mo\times\Hsp_e$, where $\Hsp_e$ is an open,
contractable subset of $\Hsp$.  Choosing $\Gsp_e$ in this way ensures
that there will exist a nowhere vanishing
$\phi\in\comega(\Gsp_e)^\heta$.  Choose one such $\phi$ and set
\begin{equation}
  \label{eq:chartoccl}
  \eta(a) = a\supl(\phi)/\phi,\quad a\in\gal.
\end{equation}
\begin{proposition}
\label{prop:indmodiso}
Equation \eqref{eq:chartoccl} defines a 1-cocycle with coefficients in
$\comega(\Mo)$ whose cohomology class is independent of the choice of
$\phi$.  Furthermore, letting $\gal$ act on $\comega(\Mo)$ by
operators $\ap+\eta(a),\;a\in\gal$ makes the map $ f\mapsto f\phi,\,
f\in\comega(\Mo)$ from $\comega(\Mo)$ to $\comega(\Gsp_e)^\heta$ into
a $\gal$-module isomoprhism .
\end{proposition}
As a consequence of this Proposition there is well defined map
$\heta\mapsto [\eta]$ from $\rH^1(\hal;\cnums))$ to
$\rH^1(\gal;\comega(\Mo))$; this map is easily seen to be linear and
injective.  The following proposition will assert that it is, in fact, an
isomorphism.  To exhibit the inverse let
$\eta\in\rZ^1(\gal;\comega(\Mo))$ be given and set
\begin{equation}
  \label{eq:ccltochar}
\heta(a) = \eta(a)(\ro),\, a\in\hal.
\end{equation}
It's not hard to check that $\heta$ annihilates $[\hal,\hal]$, and
hence can be considered as an element of $\rH^1(\hal;\cnums)$.
Furthermore, if $\eta$ is a coboundary, then $\heta=0$, and thus one
obtains a well-defined linear map from $\rH^1(\gal;\comega(\Mo))$ to
$\rH^1(\hal;\cnums)$.
\begin{proposition}
\label{prop:cformiso}
The linear map $[\eta]\mapsto \heta$, as given in \eqref{eq:ccltochar},
is an isomorphism of $\rH^1(\gal;\comega(\Mo))$ and $\rH^1(\hal;\cnums)$.
It is the inverse of the map from $\rH^1(\hal;\cnums)$ to
$\rH^1(\gal;\comega(\Mo))$ described by Proposition \ref{prop:indmodiso}
\end{proposition}
\noindent For more details, as well as an extension of this isomorphism to
higher cohomology spaces the reader is referred to \cite{Milson:coho}.

Recall that $\Lambda$ is the set of all those classes
$[\eta]\in\rH^1(\gal;\comega(\Mo))$ such that the corresponding
realization of $\gal$ by operators $\ap+\eta(a), a\in\gal$ admits a
finite dimensional invariant subspace of $\comega(\Mo)$.  One can use
Proposition \ref{prop:frobrec} (Frobenius reciprocity) to characterize
$\Lambda$. 

Before stating the theorem, two items of notation used in it have to
be explained.  First, for an $\hal$-module $U$, let
$\Lambda_\hal(U)\subset\rH^1(\hal;\cnums)$ denote the set of
$\hal$-characters that correspond to 1-dimensional submodules of $U$.
Second, for a $\gal$-module $W$, let $W^*$ denote the dual vector
space of linear forms, with the following $\gal$-action:
$$(a\cdot\alpha)(w) = \alpha(a\cdot w),\quad a\in\gal,\; w\in W,\;
\alpha\in W^*.$$ 
This gives $W^*$ the structure of a
$\gal$-antimodule, the infinitesimal analogue of a right module of a
Lie group.
\begin{theorem}
  \label{thrm:findimchar}
  Identifying $\rH^1(\hal;\cnums)$ and $\rH^1(\gal;\comega(\Mo))$ as
  per Proposition \ref{prop:cformiso}, one has
  $$\Lambda = \bigcup_W \Lambda_\hal(W^*),$$
  where the union is taken
  over all finite-dimensional, irreducible $\gal$-modules.
\end{theorem}
\begin{proof}
  Given an $\heta\in\rH^1(\hal;\cnums)$ , choose a corresponding
  $\eta\in\rZ^1(\gal;\comega(\Mo))$ as per \eqref{eq:chartoccl}.  Use
  operators $\ap+\eta(a), a\in\gal$ to give $\comega(\Mo)$ the
  structure of a $\gal$-module, and identify it with
  $\comega(\Gsp_e)^\heta$ as per Proposition \ref{prop:indmodiso}.
  Let $W$ be a finite-dimensional, irreducible $\gal$-module and note
  that a non-zero element of $\Hom_\gal(W,\comega(\Mo))$ is just a
  faithful realization of $W$ by functions on $\Mo$.  By Frobenius
  reciprocity (Proposition \ref{prop:frobrec}),
  $\Hom_\gal(W,\comega(\Mo))$ is isomorphic to
  $\Hom_\hal(W,\cnums_\heta)$.  But $\Hom_\hal(W,\cnums_\heta)$ is
  just the set of $\alpha\in W^*$ such that $\alpha(a\cdot
  w)=\heta(a)\alpha(w)$ for all $a\in\hal,\,w\in W$, i.e. $\heta$ is
  the character of a 1-dimensional $\hal$-submodule of $W^*$ spanned
  by such an $\alpha$.

  To complete the proof, note that if $\comega(\Mo)$ possesses a
  finite-dimen\-sional $\gal$-submodule, then it must certainly possess
  one that is finite dimensional and irreducible.
\end{proof}

It is important to note that Theorem \ref{thrm:findimchar} deals with
cohomology whose coefficients are $\comega(\Mo)$ rather than
$\comega(\Mo)/\cnums$, and is not, therefore, directly applicable to
the general question of quantization of cohomology.  However, as
remarked in the preceding section, if one assumes $\gal$ to be
semisimple, then
$\rH^1(\gal;\comega(\Mo)/\cnums)=\rH^1(\gal;\comega(\Mo))$, and Theorem
\ref{thrm:findimchar} becomes immediately relevant.  

In order to illustrate Frobenius reciprocity and Theorem
\ref{thrm:findimchar} it may be useful to consider the following
example.  The following three generators give a $2$-dimensional
realizations of $\fsl(2,\cnums)$ by first order differential
operators:
\begin{small}
\begin{equation}
\label{eq:cartanisoexample}
a_1=\partial_x - y^2 \partial_y ,\; 
a_2=2x\partial_x - 2y\partial_y - \lambda,\;
a_3=x^2 \partial_x - \partial_y - \lambda x,\quad
\lambda\in\cnums.
\end{equation}
\end{small}
Take $x=0,\;y=0$ to be the basepoint; it follows that the isotropy
algebra, $\hal$, is generated by $a_2$.  Using Proposition
\ref{prop:cformiso}, one sees that $\rH^1$ is $1$-dimensional, and
that $\lambda$ parameterizes the cohomology classes.  Label the
1-dimensional characters of $\hal$ by their value on $a_2$.  Let
$W_n,\,n\in\natnums$ denote the $(n+1)$-dimensional irreducible
$\fsl(2,\cnums)$ module, and note that $W_n$, as an $\hal$-module,
breaks up into $n+1$, $1$-dimensional components, $\cnums_{-n}\oplus
\cnums_{2-n} \oplus \ldots\oplus\cnums_{n-2}\oplus \cnums_n$.
Therefore by Theorem \ref{thrm:findimchar}, there exists a
finite-dimensional invariant subspace of $\comega(\Mo)$ if and only if
$\lambda$ is an integer.  Furthermore, by Frobenius reciprocity, a
copy of $W_n$ occurs in $\comega(\Mo)$ if and only if
$\lambda\in\{-n,2-n,\ldots,n-2,n\}$.  Therefore, the direct sum of all
finite-dimensional $\fsl_2$ submodules of $\comega(\Mo)$ is
\begin{equation}
\label{eqn:finmodcontent}
W_{|\lambda|} \oplus W_{|\lambda|+2} \oplus \ldots.
\end{equation}
It isn't too difficult to determine explicitly what these
finite-dimensional invariant subspaces are.  If $\lambda$ is an
integer, then each natural number $k\geq -\lambda$ indexes a
$\fsl_2$-module spanned by
$$x^i y^j (1-xy)^{-k},\quad 0\leq i\leq k+\lambda,\; 0\leq j\leq k.$$
This module is isomorphic to $W_{k+\lambda}\otimes W_k$. The module
with index $k$ is included in the module with index $k+1$, and the
totality of these modules form an infinite tower that is isomorphic to
the direct sum given in \eqref{eqn:finmodcontent}.

\section{Counter examples}
\label{sect:examples}
\begin{example}
\label{ex:counterexample1}
  This example furnishes a $3$-dimensional counter-example to
  quantization of cohomology.  Consider the following realization of
  $\fgl(2,\cnums)$: 
  \begin{align*}
    a_1&=\partial_x - y^2 \partial_y - yz \partial_z + by,
    &a_2&= x^2 \partial_x - \partial_y - rxz \partial_z - ax,\\
    a_3&= 2\partial_x - 2y \partial_y - (r+1) z \partial_z + (b-a),
    &a_4&= (r-1) z \partial_z + (a+b),
  \end{align*}
  where $x$, $y$, $z$ are local, complex coordinates, $a,b\in\cnums$
  are cocycle parameters, and $r\in\cnums$ is a realization parameter.
  Even though the above realization employs two cocycle parameters,
  $\rH^1$ is only $1$-dimensional.  Here is the reason. With $(0,0,1)$
  as the basepoint, the isotropy subalgebra is seen to be spanned by
  $(r-1)a_3+(r+1)a_4$, and hence, by Proposition \ref{prop:cformiso},
  $\rH^1$ is parameterized by $a+br$.  In this respect, note that when
  $a+br=0$, the cocycle is the coboundary of $-b\log(z)$.  Notice also
  that if both $a$ and $b$ are natural numbers, then there exists a
  finite dimensional module of functions, namely the span of
  $x^iy^j,\; 0\leq i\leq a,\, 0\leq j\leq b$.  Now if $r$ is a
  negative, irrational number, then $\{a+br:a,b\in\natnums\}$ is dense
  on the real part of $\rH^1$.  Therefore quantization of cohomology
  fails to hold for such $r$.
  
  The problem with the present example is that the isotropy subalgebra
  does not always generate a closed subgroup.  Indeed, in terms of the
  customary representation of $\fgl(2,\cnums)$ by two-by-two matrices,
  the isotropy generator is given by
  $$\begin{pmatrix} 1 & 0 \\ 0 & r \end{pmatrix}$$
  This representation makes it easy to see that if $r$ is irrational,
  then the corresponding subgroup is not closed.  However, as the following
  example will show, quantization of cohomology can fail even when
  the isotropy subgroup is closed.
\end{example}
\begin{example}
\label{ex:counterexample2}
This example will consider the homogeneous space, $\Gsp/\Hsp$, where
$\Gsp=\rSL(5,\cnums)$, and where $\Hsp$ and the corresponding Lie
algebra generator are shown below:
  \begin{equation}
  \label{eq:sl5gen}
  \begin{pmatrix}
    \E^z & 0 & 0 & 0 & 0 \\
    0 & \E^{rz} & 0 & 0 & 0\\
    0 & 0 & \E^{-(1+r)z} & 0 & 0\\
    0 & 0 & 0    & 1 & z \\
    0 & 0 & 0 & 0 & 1
  \end{pmatrix},\,z\in\cnums\qquad
    \begin{pmatrix}
    1 & 0 & 0 & 0 & 0 \\
    0 & r & 0 & 0 & 0\\
    0 & 0 & -1-r & 0 & 0\\
    0 & 0 & 0    & 0 & 1 \\
    0 & 0 & 0 & 0 & 0
  \end{pmatrix}
  \end{equation}
  It is evident that $\Hsp$ is a closed, $1$-dimensional subgroup of
  $\Gsp$.  As in the preceding example, $r$ is a parameter, and
  quantization of cohomology will fail if $r$ is an irrational number.
  It would be too cumbersome to explicitly write down the
  infinitesimal generators on the homogeneous space ($24$ generators
  in $23$ variables); it will be best to proceed abstractly.
  
  Again, label the $1$-dimensional characters of the isotropy
  subalgebra according to their value on the generator shown in
  \eqref{eq:sl5gen}.  Consider $\cnums^5$ with the obvious structure
  of an $\fsl(5,\cnums)$ module.  Restriction of $\cnums^5$ to the
  isotropy subalgebra yields $1$-dimensional characters of $\hal$ that
  are labeled by $0$, $1$, $r$, and $-1-r$.  Hence, the various tensor
  powers of $\cnums^5$ will yield characters that are labeled by
  $a+br-c(1+r)$, where $a,b,c\in\natnums$.  If $r$ is a real,
  irrational number, then the set of all such characters is clearly
  dense on the real part of $\rH^1(\hal;\cnums)$.  Therefore, by
  Theorem \ref{thrm:findimchar} quantization of cohomology fails when
  $r$ is irrational.
\end{example}
\begin{example}
\label{ex:counterexample3}
The purpose of this example is to show that quantization of cohomology
can fail even for a compact homogeneous space.  Compact, complex
homogeneous spaces were analyzed by H. C. Wang in \cite{Wang}; the
present example uses some tools from that article.  Once again take
$\Gsp=\rSL(5,\cnums)$.  For the isotropy subalgebra, $\hal$, take all
upper-triangular, nilpotent matrices, as well as the following two
generators:
  $$
a_1=
  \lp\begin{array}{ccccc}
  1 & 0 & 0 & 0 & 0 \\
  0 & -1& 0 & 0 & 0 \\
  0 & 0 & r & 0 & 0 \\
  0 & 0 & 0 &\ri-r& 0 \\
  0 & 0 & 0 & 0 &-\ri \\
  \end{array}\rp
  ,\qquad
a_2=
  \lp\begin{array}{ccccc}
  0 & 0 & 0 & 0 & 0 \\
  0 & 1 & 0 & 0 & 0 \\
  0 & 0 &-1 & 0 & 0 \\
  0 & 0 & 0 &\ri& 0 \\
  0 & 0 & 0 & 0 &-\ri \\
  \end{array}\rp
  $$
  By Proposition (7.3) of Wang's article, $\hal$ generates a
  closed subgroup of $\Gsp$.  In Proposition (3.1) Wang also shows
  that $\Gsp/\Hsp$ is compact whenever
  $$\dim\Gsp-\dim K(\Gsp) = \dim \Hsp - \dim K(\Hsp),$$
  where $K(X)$
  denotes a maximal compact connected subgroup of $X$, and $\dim$
  denotes the real dimension.  For the present case
  $K(\Gsp)=\mathrm{SU}(5)$ and $\Hsp$ has no non-trivial compact
  subgroups.  Consequently both sides are the above relation evaluate
  to $24$, and therefore $\Gsp/\Hsp$ is compact.
  
  Turning the question of quantization of cohomology, let
  $\alpha_i\in\hal^*$, $i=1,2$ denote the linear form that annihilates
  the nilpotent matrices, and satisfies $\alpha_i(a_j) = \delta_{ij}$.
  Clearly $\alpha_1$ and $\alpha_2$ are linearly independent, and span
  $\rH^1(\hal;\cnums)$.  Let $e_1,\ldots e_5$ denote the canonical
  column vector basis of $\cnums^5$.  Note that $e_1$ and $e_1\wedge
  e_2\wedge e_3$ are both eigenvectors of $\hal$ and that $\alpha_1$,
  and $r\alpha_1$ are the respective characters.  Hence, both
  $\alpha_1$ and $r\alpha_1$ are elements of $\Lambda$, and therefore
  when $r$ is a negative irrational number, $\Lambda$ is not a
  discrete subset of $\rH^1(\hal;\cnums)$.
\end{example}
\section{The semisimple case}
\label{sect:semisimple}

In the present section $\gal$ is assumed to be semisimple, and $\Gsp$
will denote the simply connected Lie group corresponding to $\gal$.
As before, $\hal$ is a subalgebra that generates a closed subgroup,
$\Hsp$, of $\Gsp$, and $\Lambda\subset\rH^1(\gal;\comega(\Mo)/\cnums)$
denotes the set of cohomology classes that admit a finite-dimensional
invariant subspace of functions.  Using the fact that
$\rH^1(\gal;\comega(\Mo)/\cnums) = \rH^1(\gal;\comega(\Mo))$ as well
as Theorem \ref{thrm:findimchar}, one can identify $\Lambda$ with the
subset of $\rH^1(\hal;\cnums)$ consisting of $1$-dimensional
$\hal$-characters obtained by restriction from finite-dimensional
$\gal$-modules.  Note that if $W_1$, $W_2$ are two $\gal$-modules and
$w_1\in W_1$, $w_2\in W_2$ are two eigenvectors of $\hal$ with weights
$\lambda_1,\,\lambda_2\in\rH^1(\hal;\cnums)$ , then $w_1\otimes w_2\in
W_1\otimes W_2$ is also an eigenvector of $\hal$ with weight
$\lambda_1+\lambda_2$.  In this way the addition operation on
$\rH^1(\hal;\cnums)$ endows $\Lambda$ with the structure of a
semigroup.  Also of interest is the abelian subgroup of
$\rH^1(\hal;\cnums)$ additively generated by $\Lambda$; it will be
denoted by $\Lambda_\hal$.  The relevance of $\Lambda_\hal$ stems from
the following fact: if $\Lambda_\hal$ is a discrete subset of
$\rH^1(\hal;\cnums)$, then the same is true of $\Lambda$.

\begin{theorem}
\label{thrm:fingen}
  $\Lambda_\hal$ is finitely generated.
\end{theorem}
\begin{proof}
  Let $\sal$ denote the radical of $\hal$.  Choose a Borel
  (maximal solvable) subalgebra $\bal\subset\gal$ that contains $\sal$.  The
  choice of $\bal$ singles out a Cartan subalgebra $\calg\subset\bal$,
  and a set of positive roots, $R^+\subset\calg^*$, such that
  $\bal=\calg\oplus\nal$, where $\nal$ is the nilpotent Lie algebra
  spanned by root vectors corresponding to $R^+$.  Let
  $\Lambda_\calg\subset\calg^*$ denote the weight lattice
  corresponding to $\calg$, and $\Lambda_\sal\subset\sal^*$ the
  abelian algebra generated by $1$-dimensional $\sal$-characters
  obtained by restriction from finite-dimensional $\gal$-modules.  Let
  $\pi_{\bal\calg}:\bal\rightarrow\calg$ denote the projection induced
  by the decomposition $\bal=\calg\oplus\nal$, and
  $\iota_{\sal\bal}:\sal\rightarrow\bal$ the inclusion injection.
  
  The first claim is that $\Lambda_\sal$ is contained in
  $\iota_{\sal\bal}^*(\pi_{\bal\calg}^*(\Lambda_\calg))$.  To that end
  let $W$ be a finite-dimensional $\gal$-module, and $w\in W$ an
  eigenvector of $\sal$ with weight $\mu\in\Lambda_\sal$.  Let
  $\Lambda_\calg(W)\subset\Lambda_\calg$ denote the set of
  $\calg$-weights that occur in the decomposition of $W$, and write
  $w=\sum_\lambda w_\lambda$, where the index runs over
  $\Lambda_\calg(W)$, and where each $w_\lambda$ is an eigenvector of
  $\calg$ with weight $\lambda$.  Choose a $\lambda_0\in
  \Lambda_\calg(W)$ such that $w_{\lambda_0}\neq 0$ and such that it
  is impossible to write $\lambda_0$ as $\lambda'+\alpha,\;
  \lambda'\in \Lambda_\calg(W)$, $\alpha\in R^+$.  Let $a\in\bal$ be
  given and write $a(w) = \sum_\lambda a(w)_\lambda$.  From the way
  $\lambda_0$ was chosen one must have $a(w)_{\lambda_0} =
  \lambda_0(\pi_{\bal\calg}(a)) w_{\lambda_0}$, and consequently
  $\mu=\iota_{\sal\bal}^*(\pi_{\bal\calg}^*(\lambda_0))$.  This proves
  the claim.
  
  Since $\Lambda_\calg$ is finitely generated, the above claim implies
  that $\Lambda_\sal$ is finitely generated too. Now
  $\rH^1(\hal;\cnums)$ is the subset of $\hal^*$ that annihilates all
  commutators of $\hal$.  Consequently, one can identify
  $\rH^1(\hal;\cnums)$ with the subspace of those elements of $\sal^*$
  that annihilate $\sal\cap[\hal,\hal]$.  In this way one can regard
  $\rH^1(\hal;\cnums)$ as a subspace of $\rH^1(\sal;\cnums)$.  It
  immediately follows that $\Lambda_\hal\subset\Lambda_\sal$, and
  therefore $\Lambda_\hal$ is finitely generated as well.
\end{proof}

Note well that $\Lambda$ is in general smaller than $\Lambda_\hal$.
This is true, for instance, in the case of the 1-dimensional
realization of $\fsl_2$ shown in the Introduction; there $\Lambda$
corresponds to all natural number values of the parameter $\lambda$,
where as $\Lambda_\hal$ corresponds to all integer values.  Even in
those instances, such as the example at the end of Section
\ref{sect:indrep} , when $\Lambda=\Lambda_\hal$, the semigroup
structure requires more generators than the group structure: at least
two generators are required to generate $\intnums$ as a semi-group,
whereas one generator suffices to generate it as a group.  This
example, as well as others known to the author, make the following
conjecture seem plausible.
\begin{conjecture}
  $\Lambda$ is finitely generated as a semigroup.
\end{conjecture}
\noindent
The proof of Theorem \ref{thrm:fingen} relied critically on the fact
that a subgroup of a free, finitely generated, abelian group is itself
free and finitely generated.  The analogous statement is not true for
semigroups.  Consequently, a proof of the above conjecture will likely
require a classification of primitive elements of $\Lambda$, i.e.  the
elements that cannot be obtained as a sum of two others in $\Lambda$.
As a starting point it seems reasonable to search for a proof of the
conjecture for the case where $\hal$ is solvable.

The fact that $\Lambda_\hal$ is finitely generated is at the heart of
the quantization of cohomology phenomenon.  When the generators of
$\Lambda_\hal$ are linearly independent (as complex vectors), then
quantization of cohomology will clearly hold.  However, the presence
of a linear relation among the generators does not automatically
destroy the quantization; in order for that to happen the relation
must be based on irrational, real coefficients.  It is upon this
principle that the counter-examples of Section \ref{sect:examples} are
constructed.  Thus, necessary conditions for quantization of
cohomology, are really just various conditions on the isotropy
subalgebra that guarantee the absence of irrational relations.  Two
instances of such necessary conditions are given below. The premise of
Theorem \ref{thrm:ratsubalg} ensures that the generators of
$\Lambda_\hal$ are linearly independent, while the premise of Theorem
\ref{thrm:cod1compact} produces a $\Lambda_\hal$ with exactly two
linearly dependent generators, but where the linear dependence is
based on a coefficient with non-zero imaginary part.

Recall that an element of a Cartan subalgebra is called rational if
the value of all roots on that element is a rational number.  A
subspace of a Cartan subalgebra will be called rational if it is
spanned by rational elements.
\begin{theorem}
\label{thrm:ratsubalg}
  If a Cartan subalgebra of $\hal$ is a rational subspace of a
  Cartan subalgebra of $\gal$, then quantization of cohomology holds.
\end{theorem}
\begin{proof}
  Choose Cartan subalgebras $\jal$ and $\calg$ of, respectively,
  $\hal$ and $\gal$, such that $\jal$ is a rational subalgebra of
  $\calg$.  Let $l$ and $m$ denote, respectively, the dimensions of
  $\calg$ and $\jal$.  Since $\jal$ is a rational subspace of $\calg$,
  one can always find a basis $a_1,\ldots a_l$ of $\calg$ such that
  the first $m$ elements span $\jal$, and such that the dual basis of
  $\calg^*$, call it $\lambda_1,\ldots,\lambda_l$, additively
  generates the weight lattice of $\calg$.  Let $\Lambda_\jal$ denote
  the abelian group additively generated by
  $\lambda_1,\ldots,\lambda_m$.  Since the $\lambda_i$ are linearly
  independent, $\Lambda_\jal$ is a discrete subset of $\jal^*$.
  
  Let $\hal=\jal\oplus\hal_1$ be the Fitting decompositions of $\hal$
  relative to $\jal$. Since $\gal$ is semisimple, $\ad_\gal(a)$ is
  diagonalizable for all $a\in\calg$, and hence the same is true of 
  $\ad_\hal(a)$ for all $a\in\jal$.  Hence, $[\jal,\hal_1] = \hal_1$,
  and a fortiori, $\hal=\jal+[\hal,\hal]$.  Hence, one can identify
  $\rH^1(\hal;\cnums)$ with the subspace of those elements of $\jal^*$
  that annihilate $[\hal,\hal]$.  With this identification one must
  have $\Lambda\subset\Lambda_\jal$, and since the latter is discrete,
  so is the former.
\end{proof}

The setting for the next theorem about quantization of cohomology will
be a compact homogeneous space.  First it will be necessary to recall
Wang's \cite{Wang} necessary and sufficient conditions on the isotropy
subalgebra $\hal$ in order that $\Gsp/\Hsp$ be compact.
\begin{theorem}[Wang's Criterion]
\label{thrm:wangcrit}
  Suppose that $\Hsp$ is connected.  Then, $\Gsp/\Hsp$ is compact if
  and only if $\hal$ satisfies the following conditions.
  \begin{enumerate}
  \renewcommand{\theenumi}{(\alph{enumi})}
  \item There exists a Cartan subalgebra, $\calg\subset\gal$, and an
    ordering of the $\calg$-root vectors such that $\hal$ is spanned
    by $\jal=\hal\cap\calg$, all the positive root vectors, and some of the
    negative root vectors.
  \item The complex $\Gsp$-subgroup generated by $\jal$
    contains a real subgroup isomorphic to $\rnums^l$, where $l$ is
    the rank of $\gal$.
  \end{enumerate}
\end{theorem}
Note that if $\jal=\calg$, then condition (a) implies that $\hal$ is a
parabolic subalgebra of $\gal$.  To understand the general case, note
that the abelian $\Gsp$-subgroup generated by $\calg$ is isomorphic
(as a real group) to $\rnums^l\times\tnums^l$.  Thus, the gist of
Wang's analysis is that in order to obtain a compact homogeneous space,
one must start with a parabolic subalgebra, and form the isotropy
subalgebra by discarding an even number of circle factors from the
Cartan subgroup.  More precisely, Wang shows that the abelian
$\Gsp$-subgroup generated by $\jal$ is isomorphic to
$\rnums^l\times\tnums^v$, where $l-v=2p$ is the number of discarded
circle factors.
\begin{theorem}
\label{thrm:cod1compact}
  Let $\Hsp$ denote the connected subgroup of $\Gsp$  generated by
  $\hal$, and suppose that $\Gsp/\Hsp$ is compact.  Let $l$ and $m$
  denote the rank of $\gal$, and $\hal$, respectively.  If $l-m\leq
  1$, then quantization of cohomology holds.
\end{theorem}
\noindent
The proof will require the following Lemma based on Wang's criterion.
Let $\calg$ and $\jal$ be as in Theorem \ref{thrm:wangcrit}.
\begin{lemma}
\label{lemma:wangcrit}
  There exists a basis $a_1,\ldots,a_l$ of $\calg$ and a basis
  $b_1,\ldots,b_{p+v}$ of $\jal$ such that 
  elements of the dual basis of $\calg^*$ additively
  generate the weight lattice of $\calg$; such that
  \begin{equation}
    \label{eq:jbasdef}
    b_i = \begin{cases}
    a_i + \sum_{j=p+1}^{2p} k_{ij} a_{j},\; 
    & i\leq p \\
    a_{p+i}, & i> p,
    \end{cases}
  \end{equation}
  where $k_{ij}\in\cnums $; and such that for every $i$ there exists
  at least one $j$ such that $k_{ij}$ is not real.
\end{lemma}
\begin{proof}
  Let $T$ denote the maximal torus contained in the subgroup generated
  by $\jal$; recall that $v=l-2p$ is its dimension.  One can always
  choose $a_1,\ldots, a_l$ such that the dual basis generates the
  weight lattice and such that $\ri a_{2p+1},\ldots,\ri a_l$ are the
  real infinitesimal generators of $T$.  Since the complex dimension
  of $\jal$ is $(l+v)/2=l-p$, one requires $p$ additional elements to
  obtain a basis of $\jal$.  Choose $\hb_1,\ldots,\hb_p$ in the span
  of $a_1,\ldots a_{2p}$ such that the $\hb_i$'s together with
  $a_{2p+1},\ldots,a_l$ are a basis of $\jal$.  Write $\hb_i = \sum_j
  \hk_{ij} a_j$ where $i$ ranges from $1$ to $p$ and $j$ ranges from
  $1$ to $2p$.  Row reduce the matrix $\{\hk_{ij}\}$ to obtain the row
  canonical form, call it $\{k_{ij}\}$, and set $b_i = \sum_j k_{ij}
  a_{j}$.  Exchanging some of the elements of the list $a_1,\ldots
  a_{2p}$, if necessary, one can without loss of generality assume
  that the $b_i$'s have the form shown in \eqref{eq:jbasdef}.
  
  Next, suppose that one of the rows, say row $i$, of $\{k_{ij}\}$
  consists entirely of real numbers.  All the elements of the row must
  be rational numbers, otherwise $b_i$ would not generate a closed
  subgroup, and consequently neither would $\jal$ nor $\hal$.
  On the other hand if the row consisted of rational numbers, then
  $\ri b_i$ would generate (in the real sense) a torus, and thereby
  violate the maximality assumption about $T$.  Therefore a
  contradiction is obtained, and the lemma is proved.
\end{proof}

\begin{proof}[Proof of Theorem \ref{thrm:cod1compact}]
  Let $\calg$, $\jal$, and their respective bases, $a_1,\ldots,a_l$
  and $b_1,\ldots,b_m$ be as in Theorem \ref{thrm:wangcrit} and Lemma
  \ref{lemma:wangcrit}.  If $l=m$ then quantization of cohomology
  holds by Theorem \ref{thrm:ratsubalg}. Suppose then that $l=m+1$.
  This implies that $p=1$, $b_1=a_1+k a_2$, where $k\in\cnums$ is not
  real, and $b_i = a_{i+1},\; i=2,\ldots m$.  Recall that the $a_i$'s
  were chosen so that the dual basis of $\calg^*$, call it
  $\lambda_1,\ldots,\lambda_l$, additively generates the weight
  lattice of $\calg$.  Let $\Lambda_\jal\subset\jal^*$ denote the
  abelian Lie algebra generated by $1$-dimensional $\jal$-characters
  obtained by restriction from finite-dimensional $\gal$-modules.
  Evidently, $\Lambda_\jal$ is generated by
  $\lambda_1,\lambda_2,\ldots,\lambda_l$.  The restrictions of
  $\lambda_3,\ldots,\lambda_l$ to $\jal$ are linearly independent,
  while $(k\lambda_1-\lambda_2)|_\jal=0$.  However, since $k$ is not
  real, $\{(n_1\lambda_1+n_2\lambda_2)|_\jal:n_1,n_2\in\intnums\}$ is
  a discrete subset of the $1$-dimensional complex vector space
  spanned by $\lambda_1|_\jal$, and therefore $\Lambda_\jal$ is a
  discrete subset of $\jal^*$.  From property (a) of Theorem
  \ref{thrm:wangcrit} one sees that $\hal=\jal+[\hal,\hal]$, and
  consequently $\rH^1(\hal;\cnums)$ can be identified with the vector
  space of those elements of $\jal^*$ that annihilate $[\hal,\hal]$.
  With this identification one must have $\Lambda\subset\Lambda_\jal$,
  and since the latter is discrete, so is the former.
\end{proof}
Example \ref{ex:counterexample3} shows that the preceding theorem is
sharp in the sense that compactness of $\Gsp/\Hsp$ alone is not
sufficient to guarantee quantization of cohomology.  Indeed, in that
example the Cartan subalgebra of $\hal$ was a codimension $2$ subspace
of a Cartan subalgebra of $\gal$.

\end{document}